\documentclass[preprint,aps,showpacs]{revtex4}
\usepackage{times}
\usepackage{amsmath}
\usepackage{amsfonts,amssymb}
\newcommand{\jgr}{J. Geophys. Res.}

\begin{document}
\title{Excitations of oscillations in loaded systems with internal 
degrees of freedom}

\author{Michael Gedalin}
\affiliation{Ben-Gurion University, Beer-Sheva 84105, Israel}
\date{\today}

\begin{abstract}
We show that oscillations are excited in a complex system under the 
influence of the external force, if the  parameters of the system 
experience rapid change due to the changes in its internal structure. 
This excitation is collision-like and does not require any phase 
coherence or periodicity. The change of the internal structure may be 
achieved by other means which may require much lower energy expenses. 
The mechanism suggests control over switching oscillations on and off 
and may be of practical use.
\end{abstract}

\maketitle

A complex system subjected to an external load is a quite common 
phenomenon in nature, laboratory, and industry. Almost all systems 
experiencing the influence of external force possess internal 
structure which is not affected by this force directly but may change 
because of intrinsic dynamics of other external causes. The examples 
of such systems span all possible scales. A planetary magnetosphere 
under the influence of the solar wind is in the equilibrium (defined 
by the stationarity of the magnetopause position)  if the 
time-independent incident 
plasma pressure is balanced by the magnetic pressure \cite{Hu95}. Yet 
the ``stiffness'' of the compressed magnetosphere is determined, 
among others,by the magnetosphere-ionosphere-atmosphere-solid earth 
interaction \cite{Yo99}. A muscle can be heuristically represented as 
a set of springs connected in parallel and in series, the elasticity 
of these ``springs'' being dependent on chemical processes and 
electrical signals in muscular fibers  \cite{Me80}. 

These systems may oscillate globally (the position of the 
magnetopause, the length of the muscle) near the equilibrium position 
(if 
not overdamped). There is a rapid growth of interest in 
possible effect of the internal variation on the excitation of global 
modes. \textcite{TU99} suggested that interaction with dynamical 
atmosphere and releases of energy due to weak local earthquakes may 
cause persisting global earth  oscillations. \textcite{KK99} find a 
relation between magnetospheric bursty bulk flows and low frequency 
Earth  
magnetic field oscillations. 
Recently, \textcite{Ch00} proposed a 
mechanism of oscillation excitation in a one-dimensional oscillatory 
system, based on the rapid change of the eigenfrequency. 
Unfortunately, their analytical consideration was inappropriate which 
obscured the physics of the phenomenon and lead them to conclusions 
which are too restricted. 

In the present paper we investigate the excitation of oscillations in 
complex systems due to rapid changes in internal parameters under 
quite general conditions. We show that such excitation (and 
oscillation amplification) is a quite common phenomenon and calculate 
its efficiency. This question may be of significant importance in 
applications. One can easily imagine systems where such excitation 
is unwanted (like building constructions). On the other hand, the 
possibility of generating oscillations by not changing external force 
but applying only weak forces (and energy) to cause local rapid 
changes of the system parameters is rather attractive. A simple 
system where both situations are possible is an LC circuit with 
distributed capacity and connected to a constant emf. Excitation of 
oscillations at the circuit frequency is unwanted when it is used in 
a 
device designed to measure electromagnetic spectra. On the other 
hand, 
such excitation would be useful in sustaining certain level of the 
current when weak damping is present and it is difficult to apply 
periodic forcing.

In what follows we consider the systems described by the following 
(vector)
equation of motion
\begin{equation}
M\ddot{\mathbf{X}}=-\frac{\partial U}{\partial \mathbf{X}} + 
\mathbf{F}, \label{eq:basic}
\end{equation}
where $\mathbf{X}$ is the vector of external parameters of the system 
(position)
which 
are subject to the external force (position of the magnetopause under 
the solar wind pressure, length of the loaded muscle, charge on the 
effective
capacitor when the LC circuit is connected to a constant emf). Let 
also $\mathbf{x}$ be the vector of internal parameters which are 
affected in a 
different way or vary at a much small time scale for some reasons. 
The 
generalized mass matrix ${M}$ and 
the potential $U$ depend both on 
the external and $\mathbf{X} $ and internal $\mathbf{x}$ coordinates, 
and the external force $\mathbf{F}$ 
depends on time. When the equations governing the behavior of the 
internal coordinates $\mathbf{x}$ are not known or are too 
complicated 
we may phenomenologically describe their influence as temporal 
dependence of $M$ and $U$.  If the system parameters and external 
force do not vary with time, 
the system is in the  equilibrium position $\mathbf{X}_{eq}$ which 
is determined by the 
condition $\partial U/\partial \mathbf{X}_{eq}=\mathbf{F}$. Near the 
equilibrium the potential can be written as 
\begin{equation}
U=\tfrac{1}{2}K
\cdot(\mathbf{X}- 
\mathbf{X}_m)\cdot ( \mathbf{X}- \mathbf{X}_m),\label{eq:eqpot}
\end{equation}
where $K$ is the stiffness 
matrix, so that one has
\begin{equation}
M\ddot {\mathbf{X}}=-K( \mathbf{X} -  \mathbf{X}_m) +
\mathbf{F}, \label{eq:near}
\end{equation}
which looks exactly as a usual oscillator equation except that now 
mass and stiffness are matrices. Recently it was shown  that the 
process of vortices 
generation in magnetohydrodynamic and shear flows 
\cite{Ch97}
can be described by a special one-dimensional case of 
Eq.~\eqref{eq:near}.

The equilibrium position is 
$\mathbf{X}_{eq}
=\mathbf{X}_m+K^{-1}\mathbf{F}$, and the general 
solution of Eq.~\eqref{eq:near} near the equilibrium point is
\begin{equation}
 \mathbf{X}= \mathbf{X}_{eq} + \sum_i a_i\hat{e}_i \sin(\omega_i 
 t+\phi_i), \label{eq:gensol}
\end{equation}
where the frequencies $\omega_i$ and unity vectors $\hat{e}_i$ are 
the 
eigenvalues and eigenvectors of the matrix $W=M^{-1}K$, respectively. 
Since the equilibrium is assumed to be stable, the matrix $W$ is 
positively determined and all frequencies are real. $a_i$ and 
$\phi_i$ 
are the amplitudes and initial phases of the normal modes. 

Mostly known channels of the energy input to the oscillating system 
include adiabatic change of the natural frequency, resonance with the 
periodically changing external force, or parametric resonance 
(see, for example, Ref.~\cite{Ar89}). The 
last two imply (quasi)periodic  behavior of the natural frequency or 
external force and are not within the scope of the present paper 
where only nonperiodic changes are considered.

The motion of the system is now fully determined by $M$, $K$, 
$\mathbf{X}_{eq}$, and $\mathbf{F}$. If these parameters change 
slowly 
at the typical time scale of oscillations, that is, the typical time 
of variation is much larger than all $1/\omega_i$, the equilibrium 
adiabatically shifts its position, while the amplitudes follow the 
well-known adiabatic law $\omega_i a_i^2\approx \text{const}$.  In 
particular, the system which was in the equilibrium in the beginning 
will remain in the equilibrium in the course of the parameter 
variation, so that no oscillations are excited.  

In the present paper we consider the case of rapid parameter changes, 
where the typical time of variation $T\ll 1/\omega_i$, for all $i$. 
We show that there this, in general, results in the 
excitation/amplification of oscillations in the system. A numerical 
example of a special type of excitation in the simplest 
one-dimensional system was recently considered by \textcite{Ch00}. 
Here we consider a most general case of nonadiabatic excitation of 
oscillations in loaded systems near the equilibrium point. 

To study quantitatively the effect let us assume that all parameters 
vary only within the time interval $0<t<T$, and $\omega_i T\ll1$. It 
is convenient to define $W=M^{-1}K$ and 
$\mathbf{f}=M^{-1}\mathbf{F}$.
From Eq.~\eqref{eq:near} one immediately has
\begin{equation}
\dot{\mathbf{X}}(T)=
\int_0^T\left[\mathbf{f}(t) -W(t) 
\left(\mathbf{X}(t)-\mathbf{X}_m(t)\right)\right], 
\label{eq:shift}
\end{equation}
where the time dependence of the system parameters is explicitly 
shown.  In what follows we denote 
the system parameters and variables at $t=0$ and at $t=T$ with 
subscripts 1 and 2, respectively. The solutions at $t\leq 0$ and 
$t\geq T$ 
will take the form:
\begin{equation}
 \mathbf{X}_k(t)= \mathbf{X}_{keq} + \sum_i 
a_{ki} \hat{e}_{ki} \sin(\omega_{ki}t+\phi_{ki}), \label{eq:before}
\end{equation}
where $k=1$ for $t<0$ and $k=2$ for $t>T$.
Since $\omega_iT\ll 1$ one can neglect the difference between 
$\mathbf{X}_1(T)$ and $\mathbf{X}_2(T)$, so that one has
\begin{equation}
\mathbf{X}_{1eq} +  \sum_i 
a_{1i} \hat{e}_{1i} \sin(\omega_{1i}T+\phi_{1i})
 =\mathbf{X}_{2eq}  + \sum_i 
a_{2i} \hat{e}_{2i} \sin(\omega_{2i}T+\phi_{2i}). 
\label{eq:fit1}
\end{equation}
The second matching condition is obtained from Eq.~\eqref{eq:shift} 
in the following form:
\begin{equation}\label{eq:fit2}
\begin{split}
& \dot{\mathbf{X}}_2(T)-\dot{\mathbf{X}}_1(T) = \int_0^T(\mathbf{f}-
\mathbf{f}_1) dt \\
&+ [\int_0^T(W-W_1)dt]\mathbf{X}_1(T) - \int_0^T 
(W\mathbf{X}_{m} - W_1\mathbf{X}_{1m})dt.
\end{split}
\end{equation}
It is convenient  to define the instantaneous equilibrium position as 
$\mathbf{X}_{eq}= \mathbf{X}_m + W^{-1}\mathbf{f}$. Taking into 
account Eq.~\eqref{eq:before} one has
\begin{equation}\label{eq:fit2a}
\begin{split}
& \sum_i\omega_{2i}a_{2i}\hat{e}_{2i} \cos(\omega_{2i}T+ \phi_{2i})=
\sum_i\omega_{1i}a_{1i}\hat{e}_{1i} \cos(\omega_{1i}T+ \phi_{1i})\\
&+ [\int_0^T(W-W_1)dt]
\sum_ia_{1i}\hat{e}_{1i} \sin(\omega_{1i}T+ \phi_{1i}) + 
[\int_0^T W(\mathbf{X}_{1eq} - \mathbf{X}_{eq})dt]. 
\end{split}
\end{equation}
Eqs.~\eqref{eq:fit1} and \eqref{eq:fit2a} allow one to find $a_{2i}$ 
and 
$\phi_{2i}$ knowing the initial state. 

It is easily seen that even when $a_{1i}=0$ for all $i$, the 
oscillation amplitude in the final state is, in general,  nonzero:
\begin{equation}
 a_{2i}^2=[\hat{e}^*_{2i}(\mathbf{X}_{2eq} - \mathbf{X}_{1eq})]^2
 +\frac{1}{\omega_{2i}^2} [\int_0^T 
\hat{e}^*_{2i}W(\mathbf{X} - \mathbf{X}_{1eq})]^2.
\label{eq:a2f}
\end{equation}
Thus, the oscillation is excited due to the (a)  irreversible shift 
of the equilibrium 
position (first term in Eq.~\eqref{eq:a2f}) and (b) reversible 
temporal shift of the equilibrium position with subsequent return to 
the same equilibrium (the second term). 
The nature of the energy input is different for the 
two mechanisms (instantaneous change of the potential energy in the 
first case, and work done by the external force because of the 
excursion of the system from the equilibrium in the second case) but 
in both cases the interaction is collision-like: certain amount of 
momentum and energy is transferred to the system in a short time 
interval. The physical nature of the excitation is quite different 
from the proposed earlier \cite{Ch00} quasi-parametric resonant interaction 
with a Fourier-component of the changing eigenfrequency. 
It is worth noting that although \eqref{eq:a2f} 
depends on the change of all parameters, $M$, $K$, and $\mathbf{X}_m$, 
the change of the mass $M$ along does not affect the instantaneous 
equilibrium position $\mathbf{X}_{eq}$ and therefore does not result 
in the oscillation excitation, as could be expected. 

Within the chosen approximation Eq.~\eqref{eq:a2f} includes all possible 
internal perturbations of $M$, $K$, and $\mathbf{X}_m$, and external 
perturbations of $\mathbf{F}$, thus giving the most general  
description for nonadiabatic excitation of complex systems. It 
includes the system considered in Ref.~\onlinecite{Ch00} as a special 
one-dimensional case where only $K$ is changed. 
It is instructive to rewrite Eq.~\eqref{eq:a2f} for this case, where
$X_{m}=0$, and $f_1=f_2$,  $W=\omega^2$, and $\omega_1=\omega_2$. 
Then the excited 
amplitude 
takes the following simple form:
\begin{equation}
a_2=\left\vert \int_0^T 
\frac{\omega^2}{\omega_1}\left(\frac{f_1}{\omega_1^2} 
- \frac{f}{\omega^2}\right)dt\right\vert, \label{eq:special}
\end{equation}
from which one can easily see that the excitation occurs always and 
not only when the eigenfrequency decreases in the perturbation (cf. 
Ref.~\onlinecite{Ch00}).

In what follows for simplicity of presentation 
we restrict ourselves with the one-dimensional case. Multidimensional 
generalization is straightforward. 
In the one-dimensional case Eqs.~\eqref{eq:fit1} and \eqref{eq:fit2a} 
immediately give
\begin{equation}\label{eq:a2}
\begin{split}
& a_2^2=[(X_{1eq}-X_{2eq})+ a_1\sin(\omega_1T+\phi_1)]^2\\
&+ \omega_2^{-2}[a_1\omega_1 \cos(\omega_1T+\phi_1) 
+(\int_0^T (\omega^2-\omega_1^2)dt) a_1\sin(\omega_1T+\phi_1)\\
&+\int_0^T \omega^2(X_{1eq}-X_{eq})dt]^2.
\end{split}
\end{equation}
In most cases the phase $\phi_1$ of the interaction beginning is 
unknown (unless the perturbation is carefully prepared with some 
definite purpose in mind). In this case one can consider the phase 
$\phi_1$ as random and average  over random distribution to obtain 
eventually:
\begin{equation}\label{eq:a2av}
\begin{split}
& a_2^2=\tfrac{1}{2}a_1^2\left[1+\frac{\omega_1^2}{\omega_2^2}
+\frac{(\int_0^T(\omega^2-\omega_1^2)dt)^2}{\omega_2^2}\right]\\ 
&+(X_{1eq}-X_{2eq})^2 + \frac{1}{\omega_2^2}\left[ \int_0^T \omega^2 
(X_{1eq} - X_{eq})dt\right]^2.
\end{split}
\end{equation}
Eq.~\eqref{eq:a2av} describes the amplification (reduction) of 
oscillations (first term) and excitation of oscillations due to the 
shift of the of the equilibrium position and to the work of the 
external force during temporal shift of the equilibrium.

In the particular case of perturbation where after time interval $T$ 
the system parameters return to their initial values, one arrives 
again at 
\begin{equation}
a_2^2-a_1^2=
\left[\int_0^T 
\frac{\omega^2}{\omega_1}\left(\frac{f_1}{\omega_1^2} 
- \frac{f}{\omega^2}\right)dt\right]^2,\label{eq:special2}
\end{equation}
where we neglected the term containing 
$[\int_0^T(\omega^2-\omega_1^2)dt]^2/\omega_2^2\ll 1$.
 Roughly speaking, during the excursion of 
the internal parameters from their equilibrium values the energy of 
oscillations increases by $\delta 
E=[\int_0^T\omega^2(X_{1eq}-X_{eq})dt]^2$ on 
average. 
If the system experiences a series of $N$ such rapid variations with 
randomly distributed phases, the 
total oscillation energy increase would be about $N\delta E$, without 
any necessity to arrange the phases or periodicity of the 
variations.   

Another efficient method of excitation is the rapid shift from the 
equilibrium position with subsequent return to this position after 
time $\gg 1/\omega$. In this case, neglecting the last term in 
Eq.~\eqref{eq:special2} and after some algebra one obtains
\begin{equation}
a^2_2=a_1^2 \left(\frac{\omega_1^2 + \omega_2^2}{2\omega_1 
\omega_2}\right)^2 + (X_{2eq}-X_{1eq})^2 \left(\frac{3}{2} + 
\frac{\omega_1^4 + \omega_2^4}{4\omega_1^2 \omega_2^2} \right), 
\label{eq:special3}
\end{equation}
which shows that this scenario always results in the oscillation 
amplification. 

To conclude,  we have shown that nonadiabatic changes in the system 
parameters and/or external force are efficient in excitation or 
amplification of oscillations in driven oscillatory systems under 
external load. 
Randomly distributed in time, 
short nonadiabatic pulses result in efficient transfer of energy to 
the system.  The energy transfer manifests itself in continuously 
increasing 
oscillation amplitude. This amplitude increase is not restricted to 
the periodically repeated (coherent) perturbations, as was suggested 
in \cite{Ch00}, but occurs in quite general conditions. 
The energy input effect is essentially nonresonant and 
more collision-like where additional momentum/energy are transferred 
to the system at the time scale much smaller than the typical 
timescale of variations in the system. 
The effect may be important for the systems whose 
natural frequencies may vary quickly due to the variable internal 
coupling. It should be emphasized that the importance of the 
above analysis is well beyond the consideration  
of simple oscillatory systems, desribed by the simple oscillatory 
potential in the form 
\eqref{eq:eqpot} (chosen here for convenience and simplicity of 
presentation),  but may be applied to quite 
general systems. The results (qualitatively) are valid for any system capable 
of (generally nonlinear) oscillations near the forced equilibrium 
position (although quantitative calculations would require knowledge 
of the structure of a particular system and ability to translate it 
into a low-dimensional description with small number of parameters). 
There is a wide spectrum of such systems, from large astrophysical 
scales (gravitationally bound systems, planetary magnetospheres under 
solar wind influence)  to usual human scale (muscles, 
constructions) and down to small scales (electric circuits). Such 
generation of oscillations may be unwanted in some systems, like 
possible excitation of internal currents in spacecraft circuits by 
cosmic rays. On the other hand,  
the essentially 
nonresonant generation of oscillations might be useful in 
experimental 
determination of the natural frequencies of the systems where it is 
difficult to apply periodic external force but where the internal 
structure can be changed relatively  easily and rapidly.  Such 
methods can be also used to re-excite damped oscillations
without changing of the main load. 

Finally, let us use  a very simple model to see whether reconnection 
at the dayside magnetopause may be responsible for excitations of 
global magnetospheric oscillations. The position of the magnetopause 
is determined by the balance of the incident plasma pressure 
$n_um_pV_u^2$ (where $n_u$ and $V_u$ are the solar wind density 
and velocity, respectively, and $m_p$  is the proton mass) and the 
magnetic pressure $B^2/8\pi$. Is the magnetopause is compressed by 
$x$, the magnetohydrodynamic conservation of magnetic flux 
\cite{Cr97} 
predicts that the magnetic field increases as $L/(L-x)$ where $L$ is 
the 
equilibrium standoff distance of the magnetopause. Thus, there 
appears 
the excess force of $\sim 2n_um_pV_u^2xA/L$, where $A$ is the 
effective area. This force has to accelerate the mass of about 
$n_dm_pAD$, where $n_d\approx 5 n_u$ is the average plasma  density,
and $D$ is the distance between 
the shock and magnetopause. Using the typical parameters $V_u\sim 
400\,$ km/s, $L\sim 10R_E$, and $D\sim R_E$, where $R_E\sim 6,000$ km 
is the Earth radius \cite{Cr97}, one finds the typical oscillation 
periods of the order of $\sim 10$ min. On the other hand, the 
typical time of reconnection should be of the order $d/v_n$, where 
$d\sim 800$ km 
is the magnetopause width \cite{Ru95}, while the velocity $v_n$ may 
be 
as high as $50$ km/s \cite{Be95}. Thus, the typical time of 
reconnection $\sim 10$ sec and much smaller than the oscillation 
period. Reconnection results in the breakdown of magnetohydrodynamics 
and therefore reduces the ``stiffness'' of the magnetic field, thus 
effectively temporarily reducing the global oscillation frequency. 
Hence, the conditions of Eq.~\eqref{eq:special2} are satisfied and 
excitation is possible. Of course, quantitative calculations require 
that we are able to translate the reconnection process into the 
change 
of internal parameters, so that at this stage the proposed scenario 
should be considered as a speculative hypothesis.


\begin{thebibliography}{}
\bibitem[Hughes(1995)]{Hu95}
W.J. Hughes,  in {\it Introduction to space physics,} (ed. M.G. 
Kivelson and C.T. Russell, Cambridge University Press, p.227,1995).
\bibitem[Yoshikava et al.(1999)]{Yo99}
A. Yoshikawa, M. Itonaga, S. Fujita, et al.
\jgr, {\bf 104, }  28437 (1999). 
\bibitem[Metcalf(1980)]{Me80}
H.J. Metcalf, {\it Topics in classical biophysics, } Prentice-Hall, 
 N. J. Englewood Cliffs, N.J., 1980.
\bibitem[Tanimoto and Um(1999)]{TU99}
T. Tanimoto and J. Um,
\jgr, {\bf 104, }  28723 (1999).
\bibitem[Kepko and Kivelson(1999)]{KK99}
 L. Kepko and M. Kivelson,
\jgr,  {\bf 104, }  25021 (1999).
\bibitem[Chagelishvili et al.(2000)]{Ch00}
G.D. Chagelishvili, A.G. Tevzadze, G.T. Gogoberidze, and J.G. 
Lominadze, Phys. Rev. Lett. {\bf 84, } 1619 (2000).
\bibitem[Chagelishvili et al.(1997)]{Ch97}
G.D. Chagelishvili,  A.G. Tevzadze, G. Bodo, and S.S. Moiseev, Phys. 
Rev. Lett., {\bf 79,} 3178 (1997).
\bibitem[Arnold(1989)]{Ar89}
V.I. Arnold, Mathematical methods of classical mechanics, (NY, 
Springer, 1989).  
\bibitem[Cravens(1997)]{Cr97}
T.E. Cravens, {\it Physics of solar system plasmas, } Cambridge 
University Press, New York, 1997.
\bibitem[Russell(1995)]{Ru95}
C.T.Russell, in {\it Physics of the magnetopause, } Geophysical 
Monograph 90, ed. P. Song, B.U.\"{O}. Sonnerup, and M.F. Thomsen, 
American Geophysical Union, Washington, 1995, p. 81.
\bibitem[Berchem et al.(1995)]{Be95}
J. Berchem, J. Raeder, and M. Ashour-Abdala, 
in {\it Physics of the magnetopause, } Geophysical 
Monograph 90, ed. P. Song, B.U.\"{O}. Sonnerup, and M.F. Thomsen, 
American Geophysical Union, Washington, 1995, p. 205.
\end{thebibliography}
\end{document}